\documentclass[twocolumn,pra,superscriptaddress,showpacs,aps,floatfix]{revtex4-1}
\usepackage{color}
\usepackage{amsmath}
\usepackage{amssymb}
\usepackage{txfonts}
\usepackage{graphicx}
\usepackage{epstopdf}
\usepackage[unicode]{hyperref}
\hypersetup{pdftitle={2D SOC BEC},
 pdfauthor={},
 pdfsubject={2D SOC BEC},
  colorlinks=true,
    linkcolor=red,
    citecolor=magenta,
    filecolor=black,
    urlcolor=blue}


\begin{document}

\title{Theoretical Origin of CP Violation in a Special FCNC-Free 2HDM with $S_N$ Symmetries}

\author{Chilong Lin}
\email{lingo@mail.nmns.edu.tw}

\affiliation{National Museum of Natural Science, 1st, Guan Chien RD., Taichung, 40453 Taiwan, ROC}

\date{Version of \today. }

\begin{abstract}
In this manuscript, a way solves the FCNC problem in 2HDM radically and leads to CP violation explicitly is demonstrated.
The derivation starts from a most general $3 \times 3$ mass matrix $M$ containing eighteen parameters, which applies to all fermion types,
and then reduces the number of parameters down to five with a Hermitian condition modified in this manuscript between the real part and imaginary part of $M$.
All matrix pairs thus derived will be diagonalized simultaneously by a same $U$ matrix which means FCNCs won't appear at tree-level anymore.
Subsequently, an assumption of $A=A_1=A_2=A_3$ is imposed and four special solutions are thus found which reveal $S_3$ and $S_2$ symmetries among fermion generations respectively.
Consequently, as a by-product of solving the FCNC problem, complex CKM matrices appear if the symmetries were assigned to up- and down-type quarks suitably.  Unfortunately, CKM elements thus derived are very different to the experimentally detected values.
The CP strength of this model, estimated by the Jarlskog invariant, is four orders higher than the one predicted by the standard model.
This difference is in fact reasonable since we don't see any such symmetries in our present universe.
However, if they had appeared in some early stages of our universe,
such a strong CP strength may provide a way to account for at least part of the discrepancy between the observed BAU and that predicted by the standard model.

\end{abstract}
\maketitle


\section{Introduction}

In the standard model (SM) of electroweak interactions CP symmetry can only be violated $explicitly$ by ranking the Yukawa couplings between fermions and Higgs fields suitably so as to bring complex phases into the Cabibbo-Kobayashi-Maskawa (CKM) matrix.
But, no one knows how these Yukawa couplings should be ranked to give such a complex CKM matrix.
In this manuscript, a way which may derive a complex, CP-violating CKM matrix explicitly will be demonstrated.
Interestingly, it is in fact a by-product of the derivation of a new Two-Higgs-Doublet Model (2HDM) free of Flavor-Changing Neutral Currents (FCNCs) at tree-level.   \\

In general, the most direct way to solve the problem of CP origin is to diagonalize the fermion-mass matrices to achieve corresponding unitary transformation matrices analytically.
However, there are too many parameters in a general $3 \times 3$ matrix to achieve the mass eigenvalues and eigenvectors analytically if no symmetries or assumptions were imposed.
Thus, an extension of SM with one extra Higgs doublet was proposed~\cite{TDLee1973}.
Through which, one expects the phases in the vacuum expectation values (VEVs) of two Higgs doublets may unlikely be rotated away simultaneously and a non-zero phase difference between them might survive so as to bring complex phases into the Lagrangian $spontaneously$.
But, this extra Higgs doublet not only failed to solve the problem of CP origin but also brought in another problem, the FCNC problem at tree-level.    \\

The FCNC problem arises when two components, $M_1$ and $M_2$ which correspond to Higgs doublets $\Phi_1$ and $\Phi_2$ respectively, of a quark-mass matrix $M=M_1+M_2$ were not diagonalized simultaneously by a same unitary transformation matrix $U$.
The non-zero off-diagonal elements in respective components $U M_1 U^{\dagger}$ and $U M_2 U^{\dagger}$ indicate flavor-changing-neutral (FCN) interactions at tree level.
The most radical way to solve this problem is to find pairs of matrices which can be diagonalized simultaneously and respectively.
But it seems also very difficult since the general texture of such matrix pairs is too complicated.       \\

At the beginning, people do not aware the FCNC problem too much.
But, as the energy and accuracy of experiments increase and no such effects were detected until 2005~\cite{Acosta2005},
the need of hypotheses to explain the smallness of such interactions emerged.
Owing to the smallness of detected such interactions, it is natural for people to consider them as contributions from loop corrections.
If we consider them as loop corrections, that means such interactions don't exist at tree level.
Thus, there must be matrix pairs which can be diagonalized simultaneously and how to find such matrix pairs becomes the key to solve this problem. \\

In 1977, Glashow and Weinberg ~\cite{Glashow1977} firstly proposed two Natural-Flavor-Conserving (NFC) models by employing a $Z_2$ discrete symmetry to forbid quarks to couple with both Higgs doublets simultaneously.
If each quark couples only to one of the Higgs doublets, surely there will be no FCNC problem.
These two models are usually referred to as the Type-I and Type-II NFC models or just Type-I and Type-II 2HDMs.
In the Type-I model, only one of the Higgs doublets couples to both quark types and the other completely does not.
In the Type-II model, up- and down-type quarks couple respectively to different Higgs doublets.
Both cases are free of FCNCs naturally since there will be only one non-vanishing component in the mass matrix of each quark type.
Besides them, there are also two similar models which are usually referred to as the Type-III and -IV models,
or sometimes the Type-Y and -X models.
However, these two models can be treated as just extensions of Type-I and -II models to include leptons~\cite[and references therein]{Aoki2009, Pich2009, Branco2012, Diaz-Cruz2004}. \\

Besides the models mentioned above,
there is another type of 2HDM which is also referred to as the Type-III 2HDM ~\cite[and references therein]{Cheng1987, Liu1987, Atwood1997, Buras2010, Sanchez2013, Crivellin2013} and usually causes confusions with the one mentioned in last paragraph.
This type of model does not deny the existence of FCNCs at tree level.
They just assume that tree-level FCNCs are highly suppressed down to the empirical values or can be canceled by loop corrections so as to fit the data.
\\

Besides the Type-I and -II NFC models mentioned above, there should be theoretically another type of NFC models.
We would like to name it a FCNC-free model to avoid confusion with those two Type-III models mentioned above.
In such a model both $M_1$ and $M_2$ are non-zero and can be diagonalized simultaneously by a same $U$ matrix.
In fact, the first such matrix pair had been discovered in a $S_3$-symmetric 2HDM ~\cite{Lin1988, Lee1990, Lin1994} decades ago.
The matrix pair achieved there were given as
\begin{eqnarray}
M_1 &=& \left( \begin{array}{ccc} A & B & B   \\  B & A & B  \\ B & B & A \end{array}\right)
={v_1 \over \sqrt{2}}\left( \begin{array}{ccc} a & b & b   \\  b & a & b  \\ b & b & a \end{array}\right), \nonumber \\
M_2 &=& i \left( \begin{array}{ccc} 0 & -D & D   \\  D & 0  & -D  \\ -D & D & 0 \end{array}\right)
=i{v_2 \over \sqrt{2}}\left( \begin{array}{ccc} 0 & -d & d   \\  d & 0 & -d  \\ -d & d & 0 \end{array}\right),
\end{eqnarray}
where ${v_1 \over \sqrt{2}}=<\Phi_1>$ and $i {v_2 \over \sqrt{2}}=<\Phi_2>$ are VEVs of Higgs doublets $\Phi_1$ and $\Phi_2$ respectively with a choice of $<\Phi_1>$ to be real.
In the coming discussions, the parameters $D$ and $d$ in our previous papers will all be replaced by $C$ and $c$ respectively for convenience. \\

But that model only solved the FCNC problem.
It still failed to give us a way to break CP symmetry since at that time we have only one such FCNC-free matrix pair for both up- and down-type quarks.
The CKM matrix thus derived was a $3 \times 3$ identity matrix $V_{CKM}=U^{(u)} U^{(d) \dagger}= {\bf 1_{3\times 3}}$ since $U^{(u)}=U^{(d)}$ in that model.
Therefore, it is natural to say that $U^{(u)} \neq U^{(d)}$ is a necessary but not sufficient condition for deriving a complex CKM matrix.
That drives us to study the problems in a less symmetric way which may probably derive different $M$ and $U$ patterns for up- and down-type quarks. \\

In this manuscript, we study the FCNC and CP problems from a very fundamental basis without any symmetries and then find $S_N$ symmetries are revealed in some special cases in which CP-violating phases are derived.
Thus, the FCNC problem is solved completely and partly the problem of CP origin.
Though CKM elements thus derived are very different to the empirical values and the CP strength predicted by this model is about four orders larger than the SM predicted one.
However, such a discrepancy could be not really harmful but rather beneficial since it can be used to account partly for the discrepancy between the cosmologically observed baryon asymmetry in the universe (BAU) and the SM predicted one.       \\

In section II, we will start the derivation from the most general $3 \times 3$ pattern of mass matrix $M=M_R + i M_I$ with eighteen parameters in it,
where $M_R$ and $i M_I$ are the real component and imaginary component of $M$, respectively.
If we assume $M$ were Hermitian, the real component and imaginary component of it will satisfy an interesting condition
\begin{equation}
M_R (i M_I)^\dagger - (i M_I) M_R^\dagger = 0,
 \end{equation}
which was firstly proposed in ~\cite{Branco1985} and will be modified and proved to be true in section II. \\

With the help of Eq.(2), the number of parameters in $M$ will be reduced from eighteen down to five.
Since five parameters are still too many for us to derive a general solution for the eigenvalues and eigenvectors of $M$,
an assumption of $A=A_1=A_2=A_3$ will be imposed and four special solutions are discovered subsequently.
It is interesting that $M$ patterns thus derived reveal $S_3$ and $S_2$ symmetries among fermion generations respectively.
One of them is the $S_3$ symmetric one in Eq.(1).  \\

As we have already four such FCNC-free matrix pairs for fermion-mass matrices,
now we have the freedom to assign different matrix pairs to up- and down-type quarks respectively and thus satisfy the necessary condition $U^{(u)} \neq U^{(d)}$ for deriving a complex CKM matrix.
Consequently, several complex CKM matrices appear in section III.
It is the first time we know how CP violation can be derived from the theoretical end. \\

Unfortunately, the CKM elements derived in section III do not coincide the empirical values very well.
The CP strengths in such models are estimated with Jarlskog invariant and found to be about four orders stronger than the value predicted by the standard model with presently detected values.
However, such a discrepancy is not strange at all since the $S_N$ symmetries do not exist in our present universe.
They may probably appeared in very early stages of our universe whose temperatures were extremely high.
If they did, the CP strength predicted here will become rational and could be used to account for part of the discrepancy between the cosmologically observed BAU and that predicted by the standard model.
Conclusions and discussions on the further developments of this derivation will be given in section IV. \\

\section{FCNC-free Patterns for Fermion-Mass Matrices}

As described above, the best way to solve the FCNC problem is to find matrix pairs whose both components can be diagonalized simultaneously.
But, there are in general eighteen parameters in a $3 \times 3$ matrix if it were complex.
It is obvious too difficult to diagonalize such a matrix analytically.
In many researches, symmetries were employed to simplify the pattern of mass matrices, or equivalently to reduce the number of parameters in them.
In ~\cite{Lin1988, Lee1990, Lin1994} a $S_3$ symmetry had been imposed among three quark generations and led to the first FCNC-free matrix pair shown in Eq.(1).
However, the $U$ matrix derived there was too simple to give a complex CKM matrix.
That hints us maybe a less symmetric matrix pattern will be better for deriving a complex CKM matrix.
Thus, we will study this topic from a completely non-symmetric beginning and then find several special solutions which are $S_N$-symmetric among generations. However, a modified Hermitian condition originally proposed in ~\cite{Branco1985} has to be discussed before the derivation since it will be very helpful. \\

If we assume the mass matrix $M=M_R+ i M_I$ were Hermitian, its real component $M_R$ and imaginary component $i M_I$ will be respectively Hermitian,
where elements in $M_R$ and $M_I$ are all real since the imaginary signs have all been extracted out from them.
For such a pair of Hermitian matrices, Eq.(2) will provide us extra conditions among their elements to reduce the number of parameters enormously.
That will be proved briefly here. \\

If there were a unitary matrix $U$ which diagonalize both $M_R$ and $i M_I$ simultaneously,
we will receive the following equations
\begin{eqnarray}
U M_R U^{\dagger} &=& U M_R^{\dagger} U^{\dagger} =M_R^{diag.}, \nonumber \\
U (i M_I) U^{\dagger} &=&  U (i M_I)^{\dagger} U^{\dagger} =M_I^{diag.},
\end{eqnarray}
where $M_R^{diag.}$ and $M_I^{diag.}$ are diagonal matrices.
Obviously, Eq.(2) is proved to be true if it were multiplied with $U$ and $U^{\dagger}$ from the left- and right-hand sides, respectively.
Thus, any matrix pairs satisfying this condition will be surely FCNC-free when applied unto the 2HDM.   \\

In the most general case, a $3 \times 3$ mass matrix can be expressed as
\begin{eqnarray}
M = M_R + i M_I &=& \left( \begin{array}{ccc} A_1 +i D_1 & B_1+ i C_1 & B_2+ i C_2   \\
 B_4 + i C_4 & A_2 +i D_2 & B_3 + i C_3  \\ B_5 + i C_5 & B_6 + i C_6 & A_3 +i D_3 \end{array}\right) \nonumber \\
 &=& \left( \begin{array}{ccc} A_1  & B_1 & B_2 \\  B_4 & A_2 & B_3 \\ B_5  & B_6  & A_3  \end{array}\right)
   + i \left( \begin{array}{ccc} D_1 & C_1 & C_2  \\  C_4 & D_2 & C_3  \\  C_5 &  C_6 &  D_3 \end{array}\right),
\end{eqnarray}
where $A$, $B$, $C$ and $D$ are all real and each of them may receive contributions from both $<\Phi_1>$ and $<\Phi_2>$ arbitrarily.    \\

It is obvious there are too many parameters in such a matrix to derive a set of analytical solutions for it.
However, if we assume $M$ were Hermitian, the $D$ parameters will be all zero and Eq.(4) becomes
\begin{eqnarray}
M &=&  \left( \begin{array}{ccc} A_1  & B_1+ i C_1 & B_2+ i C_2   \\
 B_1 - i C_1 & A_2  & B_3 + i C_3  \\ B_2 - i C_2 & B_3 - i C_3 & A_3  \end{array}\right),
\end{eqnarray}
with $B_4=B_1$, $B_5=B_2$, $B_6=B_3$, $C_4=-C_1$, $C_5=-C_2$ and $C_6=-C_3$.
Or, we may split it into
\begin{eqnarray}
M_R = \left( \begin{array}{ccc} A_1 &  B_1 &  B_2  \\  B_1 &  A_2 &  B_3  \\ B_2 & B_3 & A_3 \end{array}\right) ,  ~~~~~
i M_I = i \left( \begin{array}{ccc} 0 & C_1 & C_2   \\  -C_1 & 0 & C_3  \\ -C_2 & -C_3 & 0 \end{array}\right),
\end{eqnarray}
and then try to diagonalize them. \\

If $M_R$ and $i M_I$ were diagonalized by a same $U$ matrix simultaneously, arbitrary hybrid of them like $p M_R +i q M_I$,
where $p$ and $q$ are arbitrary complex numbers, will also be diagonalized by the same $U$ matrix.
Thus, we can allocate any two such combinations to $M_1$ and $M_2$ arbitrarily while they are still diagonalized by the same $U$ matrix.
The number of such FCNC-free matrix pairs is infinite, completely independent of the phase difference between two Higgs doublets or their VEVs.
Thus, we won't consider the allocation of them to the matrices $M_1$ and $M_2$ anymore since that seems meaningless here. \\

In Eq.(6), there are totally nine parameters and that makes the derivation still very difficult.
However, as discussed above, Eq.(2) will provide us extra conditions to reduce the number of them.
By substituting Eq.(6) into Eq.(2), we will receive
\begin{eqnarray}
 (i M_I) M_R^\dagger &=& i \left( \begin{array}{ccc} B_1 C_1 +B_2 C_2 & A_2 C_1+B_3 C_2 & B_3 C_1+A_3 C_2  \\ B_2 C_3-A_1 C_1 & B_3 C_3-B_1 C_1 &  A_3 C_3-B_2 C_1  \\ -A_1 C_2-B_1 C_3 & -B_1 C_2-A_2 C_3 &  -B_2 C_2-B_3 C_3 \end{array}\right),  \nonumber \\
 M_R (i M_I)^\dagger &=& i \left( \begin{array}{ccc} -B_1 C_1 -B_2 C_2  & A_1 C_1- B_2 C_3  &  A_1 C_2+B_1 C_3  \\ -A_2 C_1-B_3 C_2  & B_1 C_1- B_3 C_3  &  B_1 C_2+A_2 C_3  \\ -B_3 C_1-A_3 C_2  & ~ B_2 C_1-A_3 C_3  &  B_2 C_2+B_3 C_3 \end{array}\right).
\end{eqnarray}

The diagonal elements give us following conditions
\begin{equation}
B_1 C_1 =-B_2 C_2 =B_3 C_3,
\end{equation}
and the off-diagonal ones give other three
\begin{eqnarray}
(A_1-A_2) &=& ~~~(B_3 C_2+B_2 C_3)/ C_1,  \\
(A_3-A_1) &=& ~~~(B_1 C_3-B_3 C_1)/ C_2,  \\
(A_2 -A_3) &=& -(B_2 C_1+B_1 C_2)/ C_3.
\end{eqnarray}
But, substituting Eq.(8) into the sum of Eq.(9) and Eq.(10) will receive Eq.(11).
So we have in fact only four equations to reduce the independent parameters down to five.   \\

Even we have now only five parameters in a fermion type, it seems the mass matrices are still too complicated to be diagonalized analytically.
Thus, we would like to reduce the number of parameters further more by imposing some extra assumptions on them.
Here, we assume that all diagonal elements $A$ are the same, i.e.,
\begin{equation}
A_1 =A_2  = A_3=A,
\end{equation}
and we then receive relations among $B$ and $C$ parameters as
\begin{equation}
B_1^2 =B_2^2 =B_3^2,~~~~~~ C_1^2 = C_2^2 =C_3^2.
\end{equation}

By studying all possible cases one by one, Eq.(13) can only be satisfied in four cases.
It provides us a new way to solve the FCNC problem radically besides the Type-I and -II NFC models.  \\

\subsection*{Case 1: $\bold{B_1=B_2=B_3=B}$ and $\bold{C_1=-C_2=C_3=C~~~~}$}

In this case, the mass matrices can be expressed as
\begin{eqnarray}
M_R = \left( \begin{array}{ccc} A   &   B  &  B  \\  B   &  A   &  B  \\  B   &   B  &  A \end{array}\right),~~
i M_I =i \left( \begin{array}{ccc} 0  &   C  &  -C \\ -C   &  0   &  C  \\  C   & - C  &  0 \end{array}\right),
\end{eqnarray}
which are exactly the same as those derived in ~\cite{Lin1988, Lee1990, Lin1994} with a $S_3$ permutation symmetry among three fermion generations.
Subsequently, the mass eigenvalues can be analytically derived as
\begin{eqnarray}
M^{diag.} =~(m_1,m_2,m_3) = (X-Y,~ X+Y,~ Z),
\end{eqnarray}
where $X=A-B,~Y=\sqrt{3}C$ and $Z=A+2B$ were redefined to achieve a general form for the fermion mass spectra which is to appear also in other cases. \\

The $U$ matrix which diagonalize $M_R$ and $i M_I$ simultaneously is given as
\begin{eqnarray}
U_1 = \left( \begin{array}{ccc}  {{-1-i \sqrt{3}} \over {2 \sqrt{3}}} &  {{-1+i \sqrt{3}} \over {2 \sqrt{3}}} & {1\over \sqrt{3}}  \\
{{-1+i \sqrt{3}} \over {2 \sqrt{3}}} &   {{-1-i \sqrt{3}} \over {2 \sqrt{3}}} &   {1\over \sqrt{3}}  \\
{1\over \sqrt{3}} & {1\over \sqrt{3}} & {1\over \sqrt{3}}  \end{array}\right),
\end{eqnarray}
where the sub-index $k$ ($k$=1 to 4) of $U_k$ indicates to which case it corresponds. \\

As to be shown below the mass matrices can be diagonalized respectively as
\begin{eqnarray}
U_1 M_R U_1^\dagger &=& \left( \begin{array}{ccc} A-B  & 0 &  0  \\  0  & A-B &  0  \\ 0  & 0 & A+2B \end{array}\right),  \\
U_1 (i M_I) U_1^\dagger &=& \left( \begin{array}{ccc} -\sqrt{3} C & 0 &  0  \\  0  & \sqrt{3} C &  0  \\ 0  & 0 &  0   \end{array}\right).
\end{eqnarray}

Obviously, such a model is free of FCNCs at tree-level.    \\

\subsection*{Case 2: $\bold{B_1=B_2=-B_3=B}$ and $\bold{C_1=-C_2=-C_3=C}$}

In this case, the mass matrices are given as
\begin{eqnarray}
M_R =  \left( \begin{array}{ccc} A   & B  &  B  \\  B   & A  &  -B  \\ B   & -B  &  A \end{array}\right),~~
i M_I =~i \left( \begin{array}{ccc} 0   & C  & -C   \\  -C  & 0  &  -C  \\ C  & C  &  0 \end{array}\right),
\end{eqnarray}
which possesses a residual $S_2$ symmetry between the second and third generations. \\

Subsequently, the mass eigenvalues can be analytically derived as in Eq.(15) with $X=A+B,~Y=\sqrt{3} C$ and $Z=A-2B$ and the $U$ matrix thus derived is given as
\begin{eqnarray}
U_2 = \left( \begin{array}{ccc}
  {{1-i \sqrt{3}} \over {2 \sqrt{3}}} & {{-1-i \sqrt{3}} \over {2 \sqrt{3}}} &   {1\over \sqrt{3}} \\
 {{1+i \sqrt{3}} \over {2 \sqrt{3}}} &  {{-1+i \sqrt{3}} \over {2 \sqrt{3}}} & {1\over \sqrt{3}} \\
  {-1\over \sqrt{3}} & {1\over \sqrt{3}} & ~{1\over \sqrt{3}} \end{array}\right).
\end{eqnarray}

Similar to the diagonalized matrices in Eq.(17) and (18), this matrix pair is also FCNC-free at tree-level naturally.    \\

\subsection*{Case 3: $\bold{B_1=-B_2=B_3=B}$ and $\bold{C_1=C_2=C_3=C}$}

In this case, the mass matrices are given as
\begin{eqnarray}
M_R = \left( \begin{array}{ccc} A   & B  &  -B  \\  B   & A  &  B  \\ -B   & B  & ~ A \end{array}\right),~~
i M_I = i \left( \begin{array}{ccc} 0   & C  &  C   \\  -C  & 0  & C  \\ -C   & -C  & 0 \end{array}\right),
\end{eqnarray}
which possesses a residual $S_2$ symmetry between the first and third generations. \\

Subsequently, the mass eigenvalues can be analytically derived as in Eq.(15) with $X=A+B,~Y=\sqrt{3} C$ and $Z=A-2B$ and the $U$ matrix thus derived is given as
\begin{eqnarray}
U_3 = \left( \begin{array}{ccc}   {{-1+i \sqrt{3}} \over {2 \sqrt{3}}} & {{1+i \sqrt{3}} \over {2 \sqrt{3}}} &  {1\over \sqrt{3}} \\
{{-1-i \sqrt{3}} \over {2 \sqrt{3}}} &  {{1-i \sqrt{3}} \over {2 \sqrt{3}}} & {1\over \sqrt{3}} \\
  {1\over \sqrt{3}} & {-1\over \sqrt{3}} & {1\over \sqrt{3}}   \end{array}\right).
\end{eqnarray}

\subsection*{Case 4: $\bold{B_1=-B_2=-B_3=-B}$ and $\bold{C_1=C_2=-C_3=-C}$}

In this case, the mass matrices are given as
\begin{eqnarray}
M_R = \left( \begin{array}{ccc} A   & -B  &  B  \\  -B   & A  &  B  \\ B   & B  &  A \end{array}\right),~~
i M_I = i \left( \begin{array}{ccc} 0   & -C  &  -C   \\  C   & 0  & C  \\ C   & -C  & 0 \end{array}\right),
\end{eqnarray}
which possesses a residual $S_2$ symmetry between the first and second generations. \\

Subsequently, the mass eigenvalues can be analytically derived as in Eq.(15) with $X=A+B,~Y=\sqrt{3} C$ and $Z=A-2B$ and the $U$ matrix thus derived is given as
\begin{eqnarray}
U_4 = \left( \begin{array}{ccc}  {{1-i \sqrt{3}} \over {2 \sqrt{3}}} & {{1+i \sqrt{3}} \over {2 \sqrt{3}}} & {1\over \sqrt{3}} \\
{{1+i \sqrt{3}} \over {2 \sqrt{3}}} & {{1-i \sqrt{3}} \over {2 \sqrt{3}}} & {1\over \sqrt{3}} \\
 {1\over \sqrt{3}} & {-1\over \sqrt{3}} & {1\over \sqrt{3}} \end{array}\right).
\end{eqnarray}

\section{Complex CKM Matrices }

As discussed in section I, the FCNC-free $S_3$-symmetric 2HDM derived decades ago did not solve the CP problem since we had only one $U$ matrix for both quark types at that time.
That led to an identity CKM matrix in which all elements are real. \\

\begin{table}[tbp]
\centering
\begin{tabular}{|l|llll|}
\hline
$V_{CKM}$ & ~~$U_1^{(d)\dagger}$ & $U_2^{(d)\dagger}$ & $U_3^{(d)\dagger}$ & $U_4^{(d)\dagger}$ \\
\hline
~~$U^{(u)}_1$ & ~~ $1_{3\times 3}$    &   $D$                & $D^*$           &      $F$    \\
~~$U^{(u)}_2$ & ~~ $D^*$              &   $1_{3\times 3}$    & $G$             &      $E$  \\
~~$U^{(u)}_3$ & ~~ $D$                &   $G$                & $1_{3\times 3}$ &      $E^*$    \\
~~$U^{(u)}_4$ & ~~ $F$                &   $E$                & $E^*$           &      $1_{3\times 3}$ \\
\hline
\end{tabular}
\caption{\label{tab:i} Various assembles of CKM matrix.}
\end{table}

In last section we have already derived three more such FCNC-free matrix pairs with a residual $S_2$ symmetry between two of the three fermion generations.
That provides us a freedom to assign different such matrix pairs to up- and down-type quarks respectively and thus derives complex CKM matrices.
In Table.1, various $U^{(u)}$ and $U^{(d)\dagger}$ are multiplied together to see if any of them were complex.
The matrices ${\bf{1_{3\times3}}}$, ${\bf D}$, ${\bf E}$, ${\bf F}$ and ${\bf G}$ are given as follows
\begin{eqnarray}
{\bf 1_{3\times 3}} &=& \left( \begin{array}{ccc} 1 & 0 & 0  \\  0 & 1 & 0 \\ 0 & 0 & 1 \end{array}\right),
~\bold{F} = \left( \begin{array}{ccc} 2\over 3 & -1\over 3 & 2\over 3  \\  -1\over 3 & 2\over 3 & 2\over 3  \\ 2\over 3 & 2\over 3 & -1\over 3 \end{array}\right),
~\bold{G} = \left( \begin{array}{ccc} -1\over 3 & 2\over 3 & 2\over 3   \\ 2\over 3 & -1\over 3 &  2\over 3  \\ 2\over 3 & 2\over 3 & -1\over 3 \end{array}\right), \nonumber \\
\bold{D} &=& \left( \begin{array}{ccc} {{1-i \sqrt{3}}\over 3} & {1\over 3}                & {{1+i \sqrt{3}}\over 3}   \\
                                                   {1\over 3}  &  {{1+i \sqrt{3}}\over 3} & {{1-i \sqrt{3}}\over 3} \\
                                    {{1+i \sqrt{3}}\over 3}    &  {{1-i \sqrt{3}}\over 3} &  {1\over 3} \end{array}\right) =
                                    \left( \begin{array}{ccc} {2 \over 3}e^{-i {\pi \over 6}}  & {1 \over 3} & {2 \over 3}e^{i {\pi \over 6}} \\
            { 1\over 3} & {2 \over 3}e^{i {\pi \over 6}} & {2 \over 3}e^{-i {\pi \over 6}} \\
                  {2 \over 3}e^{i {\pi \over 6}} & {2 \over 3} e^{-i {\pi \over 6}} & {1 \over 3}\end{array}\right), \nonumber \\
\bold{E} &=& \left( \begin{array}{ccc} {1\over 3} &  {{1-i \sqrt{3}}\over 3} & {{1+i \sqrt{3}}\over 3}   \\
                          {{1+i \sqrt{3}}\over 3} &  {1\over 3} &  {{1-i \sqrt{3}}\over 3}  \\
                               {{1-i \sqrt{3}}\over 3} &  {{1+i \sqrt{3}}\over 3} & 1 \over 3   \end{array}\right)
                               =\left( \begin{array}{ccc} {1 \over 3} & {2 \over 3} e^{-i {\pi \over 6}} & {2 \over 3}e^{i {\pi \over 6}} \\
             {2 \over 3}e^{i {\pi \over 6}} & { 1\over 3} & {2 \over 3}e^{-i {\pi \over 6}} \\
              {2 \over 3}e^{-i {\pi \over 6}} & {2 \over 3} e^{i {\pi \over 6}} & {1 \over 3}\end{array}\right).
\end{eqnarray}

The matrices $\bold{1_{3\times 3}}$, $\bold{F}$ and $\bold{G}$ are purely real and obviously CP-conserving.
While $\bold{D}$,  $\bold{E}$ and their complex conjugates are complex, which means they are CP-violating.
It's the first time we know how CP symmetry can be broken analytically from the theoretical end.    \\

Though we have now already know the way to derive CP violation from a theoretical end.
But, amplitudes of the derived CKM elements do not fit the experimentally detected values very well.
Some of them are hundreds times higher than the detected values, say both derived $\vert V_{ub} \vert = 2/3$ in ${\bf D}$ and ${\bf E}$ are about 187 times the presently detected value $3.57 \pm 0.15 \times 10^{-3}$.
Since the CKM matrices derived in this way contain only numbers rather than any parameters.
That leaves us no space to improve the fitting between theoretical predictions and empirical values.
This could be caused by the over-simplified matrix patterns by $S_N$ symmetries.
If we can throw away the restrictions from these symmetries, maybe we can achieve some patterns which fit the empirical values better. \\

As we have already complex CKM matrices given in Table.1,
it is rational for us to go one step further to estimate the CP strength predicted by such a model.
In usual, the strength of CP violation is estimated with the dimensionless Jarlskog determinant \cite{Jarlskog1985, Tranberg2010} which was given as
\begin{eqnarray}
\Delta_{CP} &=& v^{-12} {\rm Im~ Det} [m_u m_u^{\dagger} , m_d m_d^{\dagger} ] \nonumber \\
            &=& {\it J}~ v^{-12} \prod_{\scriptstyle i < j} (\tilde{m}_{u,i}^2 - \tilde{m}_{u,j}^2 ) \prod_{\scriptstyle i < j}(\tilde{m}_{d,i}^2 - \tilde{m}_{d,j}^2 ) \simeq 10^{-19},
\end{eqnarray}
where $v$ is the Higgs vacuum expectation value and $\tilde{m}$ are particle masses. \\

The Jarlskog invariant ${\it J}$ in Eq.(26) is a phase-convention-independent measure of CP violation defined by
\begin{eqnarray}
{\rm Im} [V_{ij} V_{kl} V_{il}^* V_{kj}^* ]={\it J}~ \Sigma_{m,n} \epsilon_{ikm} \epsilon_{jln},
\end{eqnarray}
and a global fit of its value was given by the ${\bf Particle~ Data~ Group}$ as ${\it J}=(3.04^{+0.21}_{-0.20}) \times 10^{-5}$ \cite{Patrignani2016}.
Such a value corresponds to a baryon asymmetry of the universe (BAU) of the order $\eta \sim 10^{-20}$,
which is too small to account for the observed $\eta ={N_B \over N_{\gamma}} \vert_{T=3K} \sim 10^{-10}$, where $N_B$ is the number of baryons and $N_{\gamma}$ is the number of photons.
 \\

Substituting the $V_{cd}$, $V_{ts}$, $V_{cb}^*$ and $V_{us}^*$ elements of derived CKM matrices into Eq.(27),
the ${\it J}$ values for ${\bf E}$ and ${\bf E^*}$ are the same
\begin{eqnarray}
 {\it J}_{\bf E,~E^*}={16 \over 81} \sin(2\pi /3) \sim 0.171,
\end{eqnarray}
while those for ${\bf D}$ and ${\bf D^*}$ are zero since the phases in them are canceled.
The ${\it J}$ value given in Eq.(28) is almost four orders higher than the one given by the standard model.
It hints that in circumstances with $S^{ct}_2+S^{ds}_2$, $S^{ut}_2+S^{ds}_2$, $S^{uc}_2 +S^{db}_2$ and $S^{uc}_2+S^{sb}_2$ symmetries, the hyper-indices $ij$ indicate between which two generations the $S_2$ symmetry appears,
the CP strengths will be orders stronger than what we see nowadays. \\

Though the CP strengths predicted in this manuscript are much stronger than the present value.
However, it is not strange at all since such $S_N$-symmetric worlds are obviously not the one we are living in.
Surely such predictions do not fit the present values well.
If we can diagonalize the mass matrix in Eq.(5) analytically without any symmetries imposed,
which should be the status of our present universe,
the derived CKM matrix should fit the present empirical values much better than what we have done in this manuscript. \\

Besides, such a discrepancy also hint that very strong CP strength could had happened in very early stages of our universe.
Though this amount is still not enough to account for the discrepancy between the SM predicted BAU and its presently observed value.
At least, in this way, that discrepancy has been reduced for four orders. \\

\section{Discussions and Conclusions}

As demonstrated above, there is indeed a way to solve the FCNC problem in 2HDM naturally by finding matrix pairs which can be diagonalized simultaneously.
The key point of such a progress is the Hermitian condition originally proposed in \cite{Branco1985} and modified in this manuscript.
It reduces the number of parameters in a mass matrix down to five and thus enable us to find some special solutions with an assumption of $A=A_1=A_2=A_3$. \\

With such an assumption, four FCNC-free matrix pairs are discovered, including the previous $S_3$-symmetric one.
That enables us to allocate different such pairs to up- and down-type quarks respectively and thus make $U^{(u)} \neq U^{(d)}$ which is a necessary condition for deriving a complex CKM matrix.
Consequently, several complex CKM matrices appear in section III.
But, the amplitudes of CKM elements thus derived are very different to the experimentally detected values.
Even so, such a discrepancy is very rational since the CP violation derived here appears only in specific $S_N$-symmetric circumstances which are obviously not the one we are living in.
That means, in some early stages of our universe there could be epoches in which CP violation was much stronger than that in our present universe.
The Jarlskog invariant derived in section III is about four orders higher than the one predicted in the standard model with presently detected values.
That indicates the BAU we see nowadays could be legacies of some very early $S_N$-symmetric epoches of our universe.\\

Theoretically, such $S_N$ symmetries appear only in circumstances with temperatures much higher than the T=$2.73^o K$ background temperature we see nowadays.
They exist only in very early stages of our universe, maybe briefly after the spontaneous breaking of the gauge symmetry which gave masses to particles.
As shown in section III, complex CKM matrices appear only in cases that residual $S_2$ symmetries in up- and down-type quarks are between different generation pairs.
It hints the processes of breaking from $S_3$ symmetry down to $S_2$ could be different for different quark types.
In case of this, we shall see two of the three quarks in a type interact in the same way in weak interactions and such similarities appear between different generation pairs in different quark types when energy scale is high enough. \\

As the $S_N$-symmetric CKM elements derived in this manuscript are very different to the presently detected ones,
the best way to solve this problem is to diagonalize the mass matrix in Eq.(5) directly rather than imposing the assumption among $A$ parameters in Eq.(12).
That will be a world without any symmetries which is more likely the one we are living in.
If we can achieve such a solution, the fitting between them and the empirical ones shall be much better than what we have done in this manuscript.   \\

Conclusively, the FCNC problem in 2HDM is solved completely in this manuscript.
A necessary condition $U^{(u)} \neq U^{(d)}$ for deriving a complex CKM matrix is stated.
A Hermitian condition $M_R (i M_I)^\dagger - (i M_I) M_R^\dagger = 0$ is modified and proved to be true.
Besides, we also solve partly the CP problem by deriving some complex CKM matrices.
It's the first time we have a way to derive CP violation from the theoretical end.
The discrepancy between the derived CP strength and the one predicted by SM with presently detected CKM values hints the existence of very strong CP strengths in some very early stages of our universe.
That may account for at least part of the BAU we see in the present universe. \\

\end{document}